\documentclass[aps,superscriptaddress,twoside,twocolumn,floatfix,a4paper,longbibliography]{revtex4-1}

\usepackage{graphicx}
\usepackage{amsmath}
\usepackage{dcolumn}   
\usepackage{bm}        
\usepackage{amssymb}   
\usepackage{graphicx,epsfig}
\usepackage{color}
\usepackage[usenames,dvipsnames]{xcolor}
\usepackage[ocgcolorlinks,colorlinks=true,linkcolor=blue,citecolor=red]{hyperref}
\usepackage{amsmath,amssymb, amsthm, amsfonts}
\usepackage{xspace}
\usepackage{xcolor}
\usepackage{verbatim}
\usepackage{mathtools}
\usepackage[english]{babel}

\begin{document}

\selectlanguage{english}
\newcommand{\orcid}[1]{\href{https://orcid.org/#1}{\includegraphics[width=8pt]{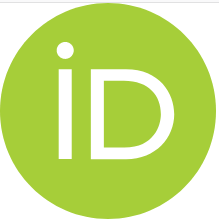}}}

\title{Preliminary Demonstration of a Persistent Josephson Phase-Slip Memory Cell with Topological Protection}

\author{Nadia Ligato \orcid{0000-0001-7655-9976}}
\altaffiliation{Present address: TeCIP Institute, Scuola Superiore Sant’Anna, Pisa, IT} 
\email{Nadia.Ligato@santannapisa.it}
    \affiliation{NEST, Istituto Nanoscienze-CNR and Scuola Normale Superiore, I-56127 Pisa, Italy}

\author{Elia Strambini\orcid{0000-0003-1135-2004}}
\email{elia.strambini@sns.it}
    \affiliation{NEST, Istituto Nanoscienze-CNR and Scuola Normale Superiore, I-56127 Pisa, Italy}

\author{Federico Paolucci\orcid{0000-0001-8354-4975}}
\email{federico.paolucci@nano.cnr.it}
     \affiliation{INFN Sezione di Pisa, Largo Bruno Pontecorvo, 3, I-56127 Pisa, Italy}
     \affiliation{NEST, Istituto Nanoscienze-CNR and Scuola Normale Superiore, I-56127 Pisa, Italy}

\author{Francesco Giazotto\orcid{0000-0002-1571-137X}}
\email{francesco.giazotto@sns.it}
    \affiliation{NEST, Istituto Nanoscienze-CNR and Scuola Normale Superiore, I-56127 Pisa, Italy}


\begin{abstract}
Superconducting computing promises enhanced computational power in both classical and quantum approaches. Yet, scalable and fast superconducting memories are not implemented.
Here, we propose a fully superconducting memory cell based on the hysteretic phase-slip transition existing in long aluminum nanowire Josephson junctions.
Embraced by a superconducting ring, the memory cell codifies the logic state in the direction of the circulating persistent current, as commonly defined in flux-based superconducting memories. 
But, unlike the latter, the hysteresis here is a consequence of the phase-slip occurring in the long weak link and associated to the topological transition of its superconducting gap. 
This disentangle our memory scheme from the large-inductance constraint, thus enabling its miniaturization.
Moreover, the strong activation energy for phase-slip nucleation provides a robust topological protection against stochastic phase-slips and magnetic-flux noise.
These properties make the Josephson phase-slip memory a promising solution for advanced superconducting classical logic architectures or flux qubits.

\end{abstract}
\maketitle
\section*{Introduction}
A Josephson junction (JJ) consists of a localized discontinuity (weak link) in the order parameter of two superconducting electrodes \cite{tinkham_introduction_2004}, where the dissipation-less current ruled by the Cooper pairs transport is controlled by the macroscopic quantum phase difference ($\varphi$) across the junction. 
Weak links are typically realized in the form of a thin insulator, a semiconductor or metallic wire, or a narrow superconducting constriction~\cite{tinkham_introduction_2004, likharev_superconducting_1979}.  
The junction current-phase relation (CPR) strongly depends on the structural attributes of the constriction, i.e., on how its effective length ($L$, i.e., the distance between the superconducting leads), width ($w$), and thickness ($t$) compare with the superconducting coherence length ($\xi\textsubscript{w}$) \cite{likharev_superconducting_1979}.
In a fully superconducting one-dimensional JJ ($w,t\ll\xi\textsubscript{w}$) the CPR  evolves from the single-valued distorted sinusoidal characteristic, typical of the short-junction limit ($L\ll\xi\textsubscript{w}$ Fig.~\ref{Fig1}a) and of non-superconducting weak links, to the multi-valued function obtained in the long regime ($L\gg\xi\textsubscript{w} $, Fig.~\ref{Fig1}b) \cite{likharev_superconducting_1979}. 
In the latter scenario, multiple (odd) solutions are available to the system at fixed $\varphi$, and the steady state will depend on the history of $\varphi$. In the specific example of Fig. \ref{Fig1}b three solutions are possible for the Josephson current ($I\textsubscript{s}$) at $\varphi$ close to $\pi$. 
Two of them are energetically-stable, they correspond to two local minima in the Josephson energy~\cite{langer_intrinsic_1967} and are topologically 
discriminated by the parity of the winding number of the superconducting phase along the wire~\cite{little_decay_1967,strambini_-squipt_2016} which reflects into two opposite directions of $I_S(\varphi)$~\cite{petkovic_deterministic_2016}, as indicated in Fig. \ref{Fig1}b by the even (red) and odd (blue) branches of $I_s$. 
In order to switch between these two stable branches, a $2 \pi$ slippage of the superconducting phase along the weak link is required. The slippage passes through the third backward solution in the CPR, a metastable state which corresponds to a saddle point in the Josephson energy separating the two stable minima and forming the barrier of a double-well potential. 
In analogy with the physics of topological insulators,
this intermediate metastable state is gapless, and is associated to the formation of a phase-slip center in the middle of the junction~\cite{langer_intrinsic_1967,arutyunov_superconductivity_2008}. 
The large superconducting condensation energy lost in this gapless center is at the origin of the strong phase-slip energy barrier separating the two topological branches.
We take advantage of this topologically-protected double well potential to implement a robust and permanent superconducting memory: The Josephson phase-slip memory (PSM). 
Differing from similar quantum phase-slip memories~\cite{mooij_phase-slip_2005}, the geometry of the PSM has been conceived for a deterministic control of the state via an external magnetic field, while stochastic quantum or thermally-activated phase slips are exponentially suppressed. As described below, these events are negligible thanks to the low resistance of the nanowire $R_N < R_q L/\xi_w$, where $R_q = h/e^2 = 6.5$~k$\Omega $\cite{virtanen_spectral_2016}. 
\section*{Results}
\subsection*{Implementation of the memory cell}
\begin{figure*}[ht!]
\includegraphics[width=0.9\linewidth]{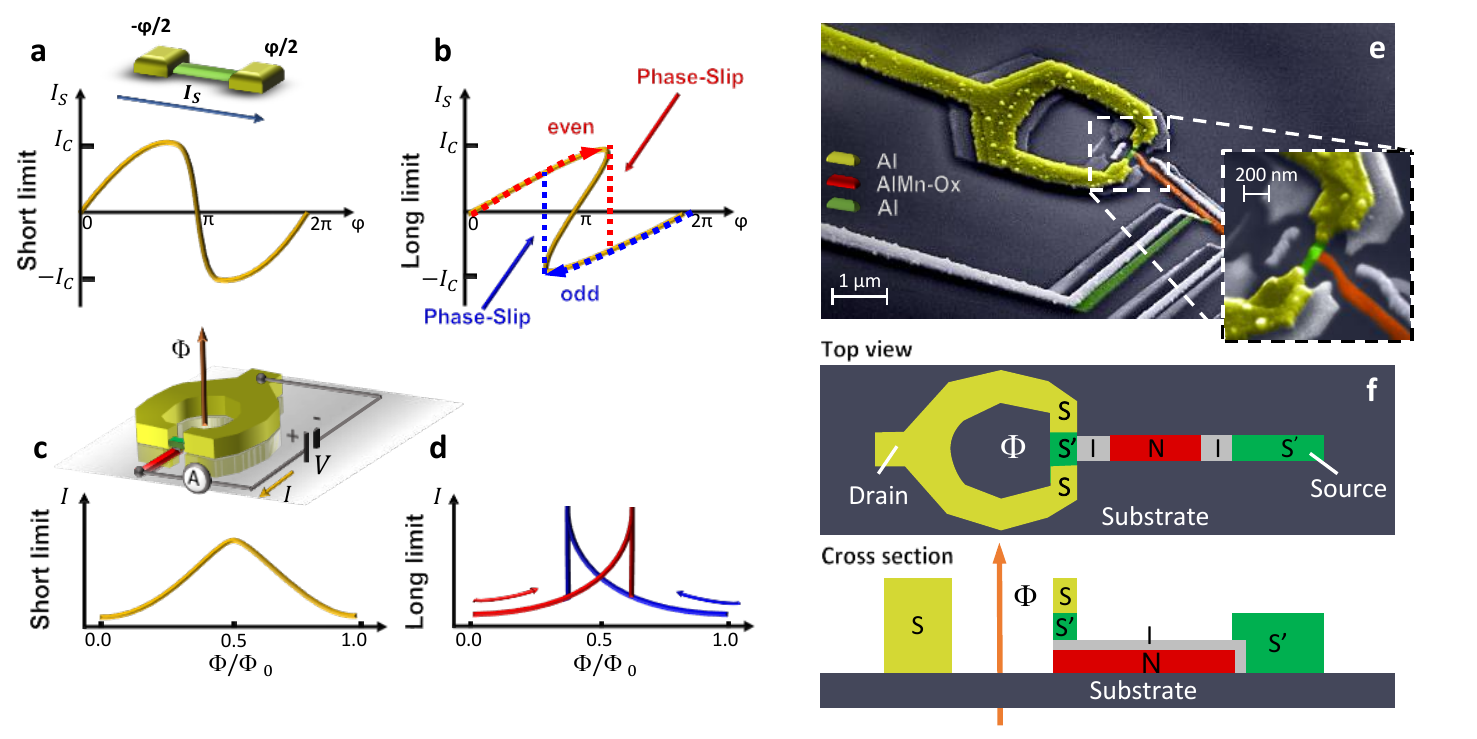} 
\caption{{Phase-Slip Memory working principle and structure.} 
\textbf{a}-\textbf{b} Sketch of the current-phase relation (CPR, $I_s(\varphi)$ ) for a S-S'-S weak link (schemed on top) in the short \textbf{a} and long \textbf{b} junction regime. 
The CPR (yellow curve) evolves from a deformed sinusoid to a multi-valued function as the junction length increases. In the latter, the transition between the two topologically-protected states, corresponding to even and odd topological index~\cite{little_decay_1967}, occurs via phase-slips in the wire~\cite{likharev_superconducting_1979,troeman_temperature_2008} and corresponds to the vertical jump indicated by the colored arrows between the two current branches (red and blue). 
\textbf{c}-\textbf{d} Dependence of the tunnel current ($I$) on the normalized applied magnetic flux ($\Phi/\Phi_0$, with $\Phi_0= h/2e \simeq 2*10^{-15}$ Wb the flux quantum), at fixed bias voltage ($V$) for a SQUIPT in the short \textbf{c} and long \textbf{d} junction regime. 
In the latter case, the current evolution shows a hysteretic profile (red and blue curves), which stems from the multi-valued CPR.
Top: scheme of a voltage-biased DC SQUIPT in a two-wire configuration. $\Phi$ is the magnetic flux piercing the ring. 
\textbf{e} Pseudo- color scanning electron micrograph of a typical PSM. An Al nanowire (green) is inserted in a micron-size Al ring (yellow), whereas an Al$\textsubscript{0.98}$Mn$\textsubscript{0.02}$ probing electrode (red) is tunnel-coupled to the middle of the nanowire and to a second Al electrode (green) via an insulating oxide layer (gray) to allow the memory operation. Inset: blow-up of the weak-link region. 
The passive replicas due to the three-angle shadow-mask metal deposition are visible.
\textbf{f} schematic top-view and cross section of the device.
}
\label{Fig1} 
\end{figure*}	
\begin{figure*}[ht!]
\includegraphics[width=0.9\linewidth]{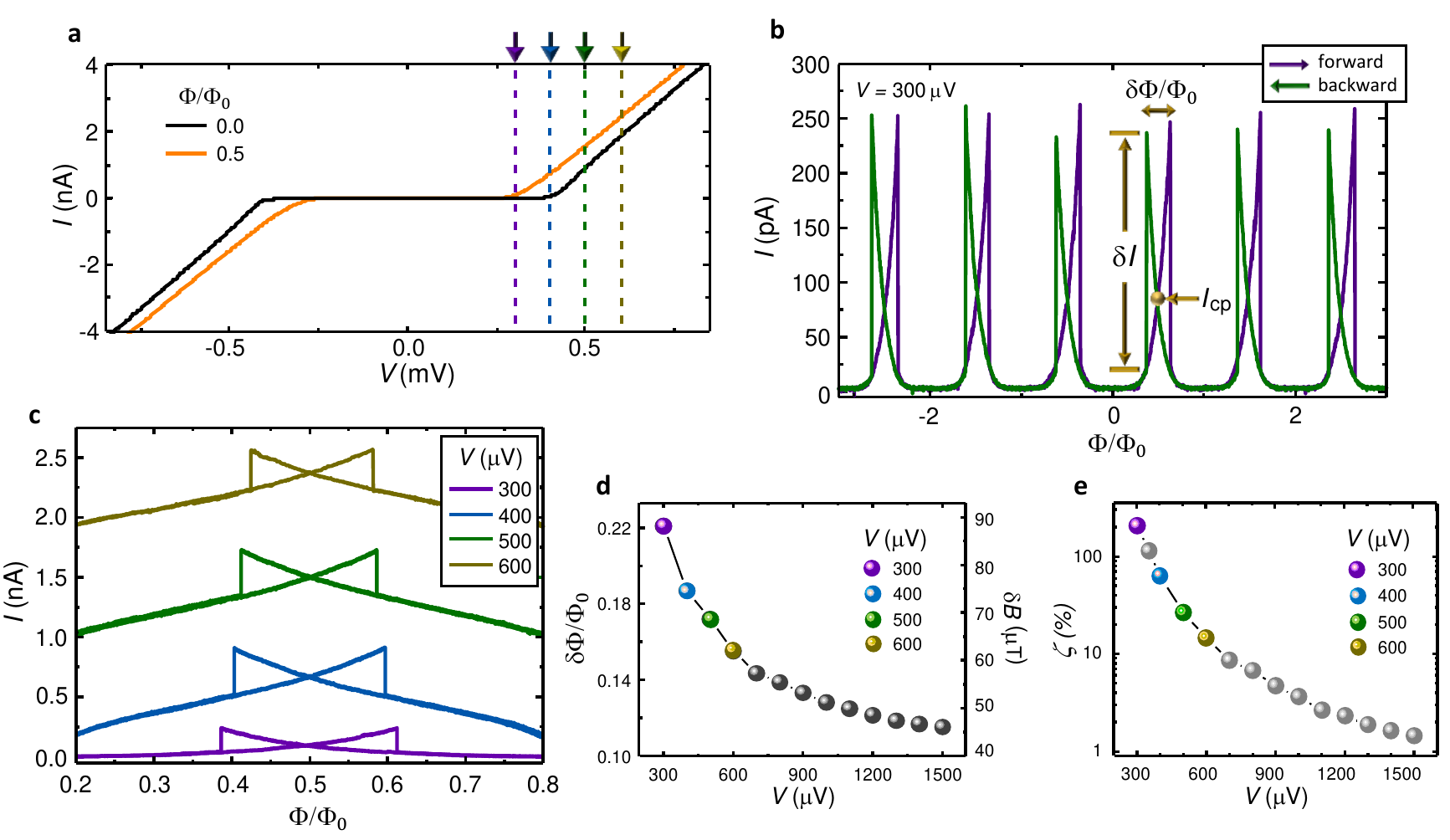}
\caption{{Phase-Slip Memory magneto-electric response.}  
\textbf{a} Current vs voltage characteristics acquired at $\Phi$ = 0 (black trace) and $\Phi$ = $\Phi\textsubscript{0}/2$ (orange trace). The magnetic flux modulates $\Delta \textsubscript{w}$ and, therefore, the  $I-V$ tunnel characteristics. 
\textbf{b} $I(\Phi)$ of a typical memory cell biased at $V$ = 300 $\mu$V. The purple and green arrows indicate the magnetic flux sweep directions. 
The width of the hysteretic loop ($\delta \Phi$), the current drop ($\delta I$), and the current at the hysteresis crossing-point  ($I\textsubscript{cp}=I(\Phi\textsubscript{0}/2)$), are also indicated. 
\textbf{c} Evolution of $I(\Phi)$ acquired for selected values of $V$, as indicated by the colored arrows in panel \textbf{a}. $I\textsubscript{cp}$ increases by rising $V$.  
\textbf{d} Dependence of the hysteresis width ($\delta \Phi$) on $V$. $\delta \Phi$ monotonically drops by increasing $V$. 
\textbf{e} Relative variation of the tunneling current ($\zeta=\delta I/I\textsubscript{cp}$) vs  $V$. All these measurements were taken at $T=25$ mK.} 
\label{Fig2}
\end{figure*}
The design of a proof-of-concept PSM requires an architecture enabling the tuning of the superconducting phase and the definition of an efficient readout scheme. To finely control $\varphi$, the JJ is inserted in a superconducting loop, where an external magnetic field gives rise to a total flux ($\Phi$) piercing the ring area. 
Stemming from fluxoid quantization~\cite{doll_experimental_1961}, 
the superconducting phase difference across the weak link is given by $\varphi=2\pi\Phi/\Phi\textsubscript{0}$ (where $\Phi\textsubscript{0} \simeq 2.067\times 10\textsuperscript{-15}$ Wb is the flux-quantum) while the phase drop along the loop is negligible (see Methods for details). The phase difference, together with the topological index, determines the amplitude of the superconducting gap in the local density of states (DOS) of the wire~\cite{virtanen_spectral_2016}, which can be probed by a metallic electrode tunnel-coupled to the middle of the junction, thereby implementing a superconducting quantum interference proximity transistor (SQUIPT)~\cite{giazotto_superconducting_2010}, as sketched on top of Fig.~\ref{Fig1}c.
As a result, at fixed $\Phi$ the amplitude of the tunneling current ($I$) flowing through the probing electrode will depend on the even/odd parity of the topological index of the junction codifying the logic [0] and [1] states of the PSM cell (see Fig. \ref{Fig1}d).
Encoding the memory state in the parity of the winding number is a common feature to all flux-based superconducting memories, including e.g. nano-SQUIDs\cite{murphy_nanoscale_2017,ilin_supercurrent-controlled_2021} flux qubits\cite{mooij_superconducting_2006} or kinetic-inductance memories \cite{chen_kinetic_1992} from which it shares the low dissipation and high operation speeds.
But, differing form the latter approaches, the dynamics of the memory cell here is entirely dominated by the physics of the weak-link. The read out in the SQUPIT is based on a tunneling spectroscopy of the weak-link and the hysteresis in the magnetic flux is not a consequence of an unbalance between the ring and junction inductance but is an intrinsic property of the CPR.

The scanning electron micrograph (SEM) of a representative PSM cell is shown in Fig.\ref{Fig1}e together with a top-down and cross-section scheme in Fig.\ref{Fig1}f. Realized through a suspended-mask lithography technique (see Methods for fabrication details) the weak link consists of a one-dimensional Al nanowire (green, $t$ = 25 nm and $w$ = 90 nm) with a length $L$ $\sim{400}$ nm, embedded in a micron-sized 70-nm-thick Al ring (yellow). In addition, a 20-nm-thick normal metal electrode (red, Al$\textsubscript{0.98}$Mn$\textsubscript{0.02}$) is tunnel-coupled to the center of the wire (with a normal-state tunnel resistance $R_{t1}\simeq 65$ k$\Omega$).
To measure the tunneling current, a second Al lead (green) is tunnel-coupled to the normal metal electrode (with a normal-state resistance $R_{t2}\simeq 90$ k$\Omega$)~\cite{ronzani_phase-driven_2017}. 
Based on the device structural parameters, we estimate the ratio $L/\xi\textsubscript{w,0}\simeq 6$, where $\xi\textsubscript{w,0}\simeq65$ nm is the zero-temperature coherence length\cite{de_simoni_metallic_2018}, thereby providing the frame of the long-junction regime~\cite{likharev_superconducting_1979,virtanen_spectral_2016} (see Methods for details). 
Within these geometrical constrains and thanks to the low resistivity of Al ($\rho < R_q \xi_w$), both quantum and thermally-activated phase slips are negligibly small, with rates $< 10^{-289} $Hz (see Methods for more details on the estimate). 
Notably, the PSM is completely made of aluminum compounds thus ensuring high-quality tunnel barriers and full compatibility of all fabrication steps for industrial scaling.

\subsection*{Magneto-electric response}

To test the PSM transport properties and assess the operation parameters of the memory cell, we first performed a preliminary magneto-electric characterization at bath temperature $T=25$ mK. Figure \ref{Fig2}a shows the current vs voltage characteristics ($I(V)$) of a typical device measured at $\Phi= 0$ (black curve) and  $\Phi= \Phi\textsubscript{0}/2$ (orange curve). 
At zero magnetic flux, the quasiparticle tunnel current is suppressed for $|V| \lesssim  400\;\mu$V due to the presence of two S-I-N tunnel junctions in series and is consistent with the an Al gap of $\simeq 200\; \mu$eV for both  the read-out lead ($\Delta \textsubscript{Al} $) and the weak link ($\Delta \textsubscript{w}(\Phi=0)$).
The latter can be modulated by the external magnetic flux~\cite{giazotto_superconducting_2010,ronzani_phase-driven_2017}, showing a reduction of about $50\%$ at $\Phi= \Phi\textsubscript{0}/2$ (orange line), $\Delta \textsubscript{w}(\Phi=\Phi_0/2) \simeq 100\; \mu$eV (see also Supplementary Figure~1 for more details).
Differently from short-junction SQUIPTs~\cite{ligato_high_2017,ronzani_phase-driven_2017},
the $I(\Phi)$ characteristic is not only $\Phi\textsubscript{0}$-periodic, but it is also strongly hysteretic in $\Phi$.
This is highlighted in Fig. \ref{Fig2}b, where the tunnel current measured at $V = 300\; \mu$V as a function of increasing (purple trace) and decreasing (green trace) magnetic flux is shown. 
The forward trace exhibits periodic maxima followed by sudden jumps corresponding to the nucleation of a phase-slip center in the superconducting nanowire~\cite{likharev_superconducting_1979,virtanen_spectral_2016,arutyunov_superconductivity_2008}.
Accordingly, the backward trace evolves in a totally specular fashion. 
The evolution of $I(\Phi)$ on the bias voltage is shown in Fig. \ref{Fig2}c. 
The hysteresis loop drawn by the back and forth $I(\Phi)$ exhibits a reduction of its width ($\delta\Phi$) by increasing $V$, as quantified also in Fig. \ref{Fig2}d. 
This trend can be ascribed to a local overheating in the weak link induced by the quasiparticle current flowing through the probing junction which
enlarges $\xi\textsubscript{w}(T)$~\cite{tinkham_introduction_2004} thereby deviating the CPR towards the single-valued non-hysteretic form \cite{likharev_superconducting_1979,virtanen_spectral_2016}.
The relative separation between the two $I(\Phi)$ branches can be quantified by a parameter ($\zeta$) defined as the ratio between the current drop at the phase-slip transition and the current at the hysteresis crossing point, $\zeta=\delta I/I(\Phi=n\Phi_0/2)$, where n is an integer odd number. A large $\zeta$ improves the visibility of the PSM logic states. 
Similarly to $\delta\Phi$, the increase of $V$ induces a monotonic reduction of $\zeta$, as shown in Fig.~\ref{Fig2}e.

\begin{figure}
\includegraphics[width=\linewidth]{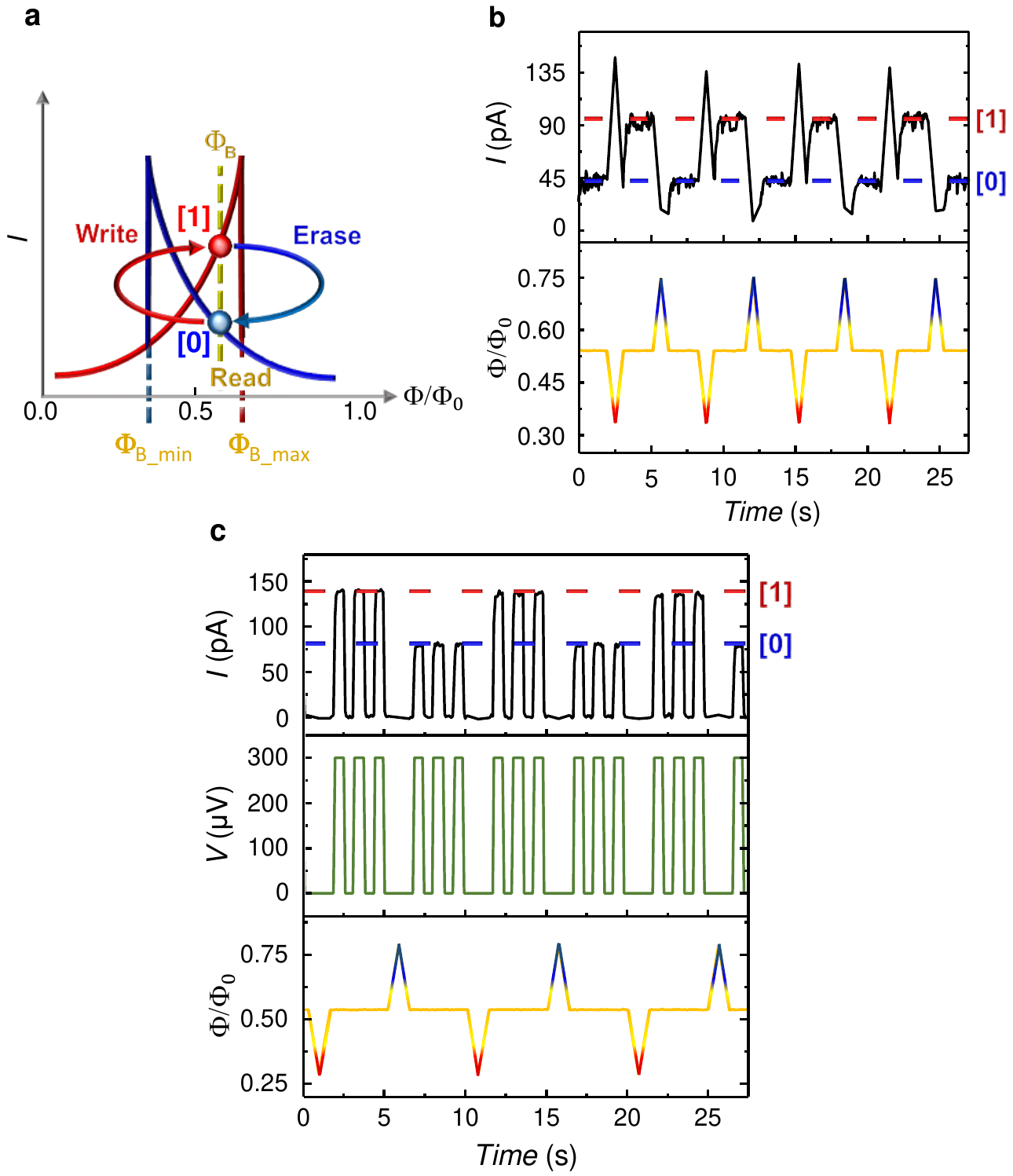}
	\caption{{Operation of the Phase-Slip Memory with DC read-out.} 
	\textbf{a} Sketch of the memory operation principle at a constant voltage bias ($V$). Low (blue, $I\textsubscript{[0]}$) and high (red, $I\textsubscript{[1]}$) current branches at the bias flux ($\Phi\textsubscript{B} \in (\Phi\textsubscript{0}/2,\Phi\textsubscript{B\textunderscore max})$) encode the [0] and [1] logic states, respectively. 
	The erase (write) operation is performed by applying a flux pulse with amplitude $\Phi\textsubscript{E}>\Phi\textsubscript{B\textunderscore max}$ ($\Phi\textsubscript{W}<\Phi\textsubscript{B\textunderscore min}$). 
	The memory can also be operated in the complementary part of the hysteresis at  $\Phi\textsubscript{B} \in (\Phi\textsubscript{B\textunderscore min},\Phi\textsubscript{0}/2)$ by exchanging the erase and write fluxes. 
	\textbf{b} Evolution of the read-out tunneling current (top panel) measured at $V=300\;\mu$V for $\Phi$ composed by a bias flux $\Phi\textsubscript{B}=0.54\Phi\textsubscript{0}$ (yellow trace) interrupted by write ($\Phi\textsubscript{W}=0.33\Phi\textsubscript{0}$, red) and erase ($\Phi\textsubscript{E}=0.75\Phi\textsubscript{0}$, blue) pulses (bottom panel).     
	\textbf{c} Same as in \textbf{b} but now the voltage bias (central panel, green trace) is applied only during the read-out operation to minimize power consumption and demonstrate the non-volatility of the memory cell.
	All the measurements were taken at $T=25$ mK.
	}
\label{Fig3}
\end{figure} 
\begin{figure*}[ht!]
\includegraphics[width=0.9\linewidth]{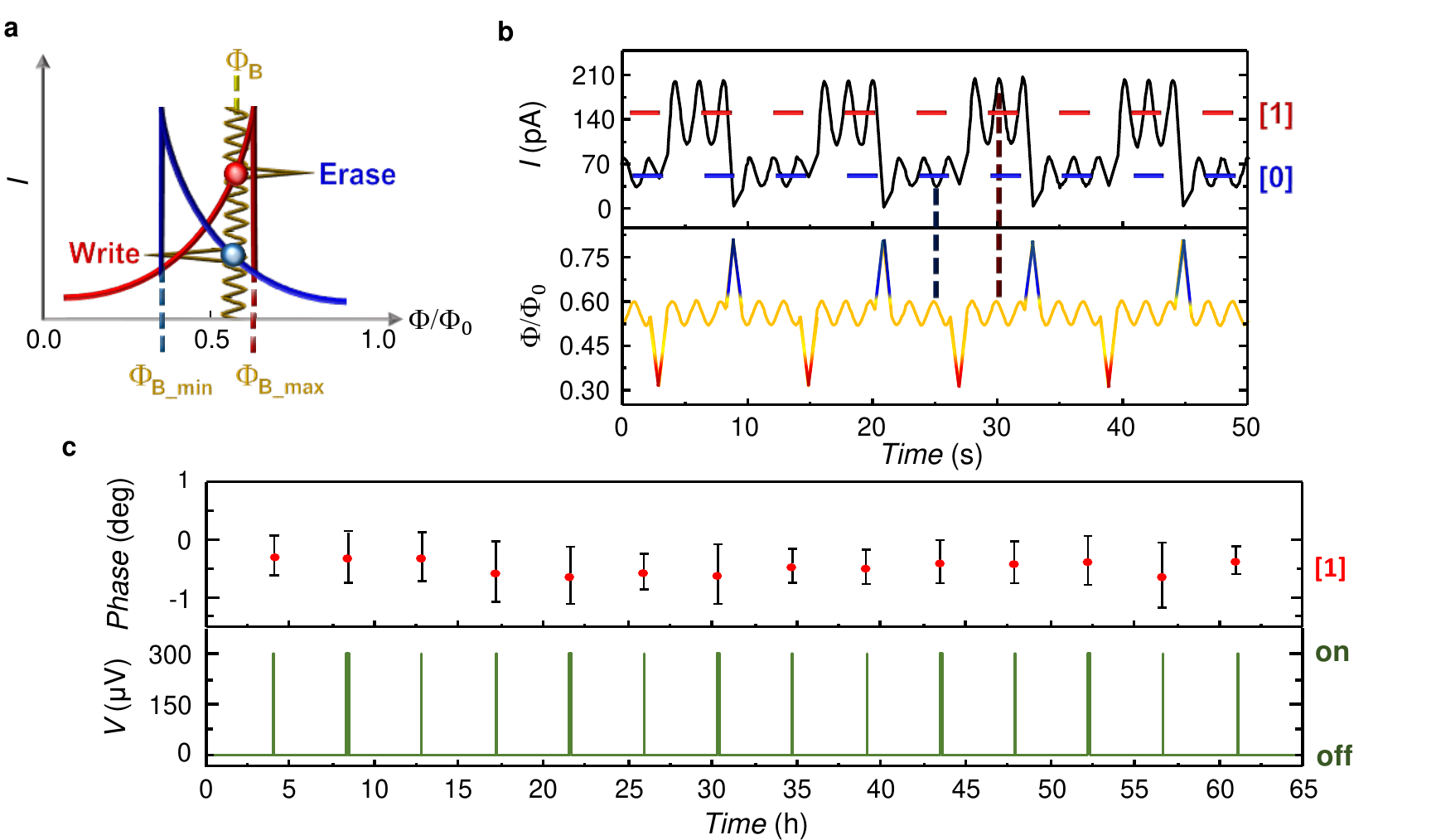}
	\caption{{Operation of the Phase-Slip Memory with AC read-out.} 
	\textbf{a} Sketch of memory operation in the presence of a sinusoidal flux oscillation ($\Phi\textsubscript{AC}$, yellow trace) around $\Phi\textsubscript{B} \in (\Phi\textsubscript{0}/2,\Phi\textsubscript{B\textunderscore max})$. 
	\textbf{b} Evolution of the read-out current (top panel) measured at $V=300 \; \mu$V and $\Phi$ composed by a flux bias ($\Phi\textsubscript{B}=0.56\Phi\textsubscript{0}$) superimposed with a sinusoidal oscillation $\Phi\textsubscript{AC}=\pm0.04\Phi\textsubscript{0}$ (yellow trace in the bottom panel). Write ($\Phi\textsubscript{W}=0.32\Phi\textsubscript{0}$, red) and erase ($\Phi\textsubscript{E}=0.81\Phi\textsubscript{0}$, blue) flux pulses are applied to switch the logic state of the memory cell.
	Notice that the two current signals oscillate with a $\pi$ shift making the phase of the AC signal a very sensitive read-out observable. Vertical dashed lines highlight the signals phase shift with respect to the magnetic flux. \textbf{c} Demonstration of persistent memory operation at  $\Phi\textsubscript{\textbf{}}=\Phi\textsubscript{0}/2$ obtained by measuring the signal phase with a lock-in amplifier (top) every 4 hours and only when the read-out voltage is turned on ($V=300 \; \mu$V, bottom).
	State [1] was measured for almost 3 days showing no sign of degradation, and low dissipation being $V=0$ for most of the time. The error bar was estimated from the root mean square of the sampled signal.   
	All the data were recorded at $T=25$ mK.}
\label{Fig4}
\end{figure*} 
\subsection*{Memory operation with DC readout}

The typical operation cycle of the PSM memory cell is sketched in Fig.~\ref{Fig3}a.
A bias flux ($\Phi_B$) is required to access the multi valued state enclosed within the hysteretic domain ($\Phi\textsubscript{B\textunderscore min} = (\Phi_0-\delta\Phi)/2  ,\Phi\textsubscript{B\textunderscore max}=(\Phi_0+\delta\Phi)/2$).
Writing (erasing) operations are performed by lowering (increasing) the total flux below (above) the hysteretic domain by means of short pulses. As a consequence, the parity of the topological index switches between odd and even and the tunneling current between low and high current state. 
Figure \ref{Fig3}b shows a real-time writing/erasing operation in the continuous read-mode, i.e., with a fixed a bias voltage $V=300\;\mu$V. The bias flux is set at $0.54\Phi_0$, just above the crossing-point of the hysteresis to avoid degeneracy in the current amplitude (see Fig. \ref{Fig2}c). The memory is then initialized in the [0] state corresponding to a current $I\simeq 43$ pA. By applying a negative flux pulse down to $\Phi\textsubscript{W}=0.33\Phi\textsubscript{0}$, the PSM logic state suddenly transits to [1] as detected by the current jump to $I\simeq 90$ pA. Conversely, the logic state [0] is recovered via a positive erasing flux pulse up to $\Phi\textsubscript{E}=0.75\Phi\textsubscript{0}$. The device unequivocally shows the typical behavior of a memory cell upon many erasing/writing cycles.
From the real-time characteristic is possible also to quantify the energy required for the writing/erasing operations. This can be estimated from the energy difference of the system in the two flux configurations that can be simplified in $E(\Phi_{B_{max},B_{min}})-E(\Phi_{0})\simeq \frac{\Phi_0}{2 \mathcal{L}_{K}} \frac{\delta\Phi}{2}$, where $\mathcal{L}_{K}$ is the kinetic inductance of the JJ \cite{mooij_phase-slip_2005}. 
In our experimental configuration, the estimated energy is $\sim 0.1$~eV, this number is consistent with the predictions for the energy of the topological barrier $U \sim \Delta \textsubscript{w} \frac{\hbar}{e^2 R_N} \frac{L}{\xi_w}$~\cite{virtanen_spectral_2016}. 
Notably, differing from conventional flux-based superconducting memories, the inductance of the PSM ring is not relevant for the device which can be made negligibly small without any loss of hysteresis or functionality. This allows the miniaturization of the PSM that could be further operated with a flux generated by supercurrents directly injected in a small portion of the superconducting ring~\cite{enrico_-chip_2019} therefore eliminating the requirement of an external magnetic field but with the disadvantage of an additional feed line integrated in the device.

The ability of a memory cell to retain the data even when the power is temporarily turned off is called non-volatility, which, even if not essential for a RAM memory, it is an adding value for energy saving and data storage. 
The PSM requires two power sources: one to generate the bias flux $\Phi_B$ and one for the read-out signal. 
The former was provided by an external superconducting magnetic controlled by a current source, then power dependent. To overcome this limitation $\Phi_B$ could also be generated by a permanent dissipationless superconducting coil as well as a metallic ferromagnetic layer buried in the semiconducting substrate or by directly employing a ferromagnetic insulator as dielectric substrate\cite{strambini_revealing_2017,de_simoni_toward_2018}. Alternatively, a proper phase bias might be generated with an additional ferromagnetic pi-junction~\cite{ryazanov_coupling_2001} inserted in the ring or through a phase-battery~\cite{strambini_josephson_2020}.
The read-out voltage is only required to probe the resistance state of the PSM. As demonstrated in Figure \ref{Fig3}c, temporarily and repeated measures of both logic states do not affect the stored data with a readout dissipation as low as $P_{[0]}\simeq25$ fW and $P_{[1]}\simeq40$ fW for logic state [0] and [1], respectively, and only limited by the noise of the current amplifier.
This low dissipated power combined with the intrinsic cutoff time $\tau_{R}\simeq 30$ ps estimated from the RC circuit of the tunnel junctions (see Methods for details) yields a predicted tiny energy required per bit readout
$J_{[0]}=P_{[0]}\tau_{R}\simeq 4.7$~$ \mu$eV and $J_{[1]}=P_{[1]}\tau_{R}\simeq 7.5 $~$ \mu$eV.
These values were only estimated, and stem from the severe bandwidth limitations of the cryogenic filters. 
Similarly to rapid single flux quantum, the writing/erasing process is expected with a switching time of $\sim 1$~ps which is typical for small superconducting loops~\cite{golod_single_2015,zhao_compact_2018,ryazanov_magnetic_2012}.
The PSM speed is therefore expected to be on par with current state-of-the-art superconducting memories both in the reading and in the writing/erasing process~\cite{vernik_magnetic_2013,gingrich_controllable_2016,golod_single_2015,madden_phase_2018,zhao_compact_2018}.

\subsection*{Memory robustness and operation with AC readout}

\begin{figure}[ht]
\includegraphics[width=\linewidth]{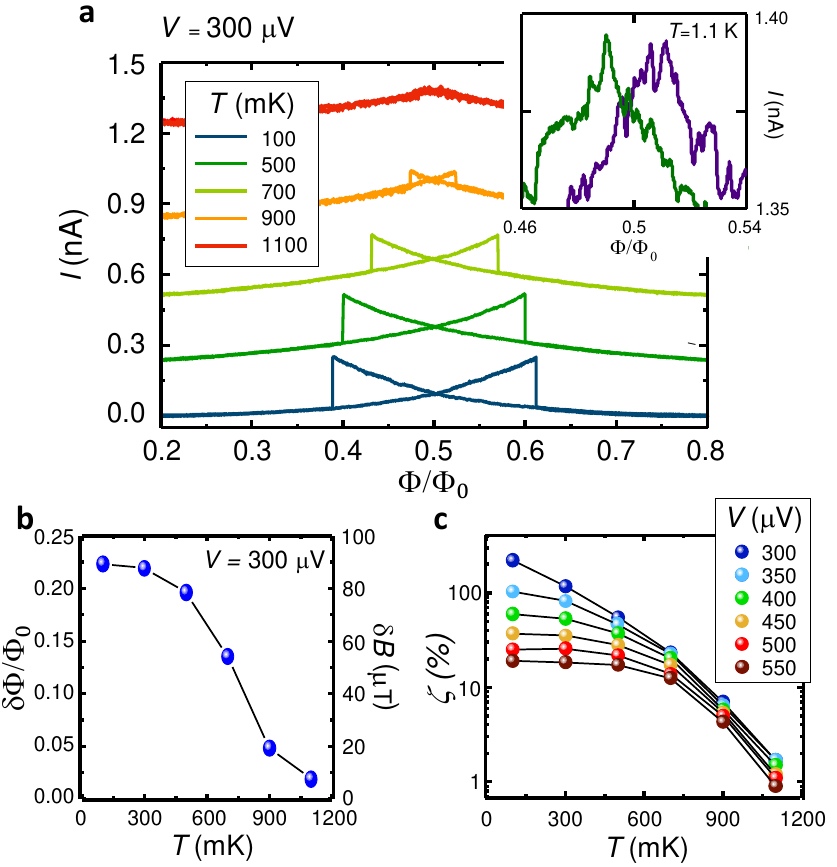} 
\caption{{Temperature dependence of the Phase-Slip Memory.} 
\textbf{a} Current modulation $I(\Phi)$ for several bath temperatures ($T$) at $V= 300$  $\mu$V. 
The hysteresis loop narrows and fades out by increasing the temperature since the superconducting nanowire approaches the short-junction limit at high $T$. Inset: blow up of the $I(\Phi)$ characteristics around $\Phi_0/2$ at $1.1$ K. Forward (purple) and backward (green) traces highlight the presence of hysteresis. 
\textbf{b} Temperature dependence of $\delta\Phi$ measured at $V= 300$ $\mu$V.
$\delta\Phi$ monotonically decreases with temperature. 
\textbf{c} $\zeta$ vs $T$ for selected values of $V$.  $\zeta$ drops with temperature, and by increasing $V$. Black lines in panels b and c are guides for the eye.}
\label{Fig5}
\end{figure}

The robustness of the PSM against flux fluctuations is tested by superimposing to the working biasing flux a sizable sinusoidal signal ($\Phi\textsubscript{AC}$, see Fig. \ref{Fig4}a).
The PSM shows optimal stability with respect to flux oscillations, as shown in Fig. \ref{Fig4}b for $V=300\;\mu$V and $\Phi\textsubscript{B}=0.56\Phi\textsubscript{0}$. 
The memory preserves the stored state and keeps the readout value of the two logic states well separated for fluctuations  $\Phi\textsubscript{AC}\simeq 0.08\Phi\textsubscript{0}$,  then $\sim 50\%$ of the hysteretic domain of the memory $\delta \Phi$, at least.
Interestingly, thanks to the opposite sign of the magnetoconductance of PSM in the two topological states (visible for instance in Fig.~\ref{Fig2}b and c), the AC flux modulation induces an AC response in the tunneling current which acquires a $\pi$ shift when switching between the two logic states [0] and [1]. 
This phase shift provides a complementary and efficient method to probe the parity of the JJ winding number, which is not affected by the position of $\Phi_B$ within the hysteretic domain, or by the low visibility of the DC readout signal (see also Supplementary Figure~4 and 5 for more details). 
This allows to operate the memory cell also in the degenerate point $\Phi_B = \Phi_0/2$, where the energies of the [0] and [1] states are equal, a basic condition to implement a phase-slip qubit~\cite{mooij_phase-slip_2005,mooij_superconducting_2006}. Therefore, the PSM provides an alternative low-frequency method for the qubit readout.
With the phase-based readout the persistency of the PSM have been tested up to almost three days, as shown in Figure \ref{Fig4}c.
The memory is initialized to logic state [1], and the readout is performed every 4 hours. 
No sign of signal degradation has been observed even after $\sim 3$ days of measurement confirming the vanishing phase-slip rate ($\sim 10^{-289}$Hz) as estimated from our parameters~\cite{virtanen_spectral_2016,arutyunov_superconductivity_2008} (See Methods for details on the estimate). 
As a consequence, the memory error rate expected for quantum and thermally-activated phase slips is infinitesimally small and errors can be generated only by large magnetic-flux fluctuations ($\gtrsim \delta \Phi$) of the driving magnetic flux.
The other source of error that might degrade the memory state is the reading current that could switch the memory via inductive coupling to the ring or by quenching the superconductivity of the weak-link, as commonly happen for superconducting kinetic inductance memories\cite{ilin_supercurrent-controlled_2021}. Differing from the latter, the high resistance of the probing tunnel barrier strongly limits the reading current to $\lesssim$nA, then much smaller than the current required for switching ($\sim$mA)\cite{enrico_-chip_2019} and the critical current of the weak-link ($\gtrsim \mu$A for an Al nanowire\cite{bours_unveiling_2020}). This makes also the error rate during readout operation negligible. 
High temperature can degrade the performance of PSM by increasing $\xi_w(T)$~\cite{tinkham_introduction_2004} thereby lowering the JJ effective length, and driving the nanowire junction towards the non-hysteretic single-valued CPR occurring for $L\lesssim3.5\xi\textsubscript{w}$~\cite{likharev_superconducting_1979,troeman_temperature_2008}. 
In addition, thermal activation can substantially increase the phase-slip rate in the vicinity of the transition that is at $\phi \lesssim \phi_{B_{max}}$ and $\phi \gtrsim \phi_{B_{min}}$)~\cite{virtanen_spectral_2016}.
Figure \ref{Fig5}a shows the evolution of the hysteresis loop at several bath temperatures ($T$). The hysteresis progressively fades out by increasing $T$, but persists up to $1.1$ K, which corresponds to $\sim 85\%$ of the nanowire critical temperature, with $\delta\Phi$ reduced to the $\sim12\%$ of the base temperature value (see Fig. \ref{Fig5}b).
Consequently, also the contrast $\zeta(T)$ lowers by increasing $T$, as shown in Fig.~\ref{Fig5}c.
Still, the visibility of the hysteresis loop at high-temperatures demonstrates the strength of the PSM with a substantial protection of the topological state even in the presence of a sizable amount of hot quasi-particles~\cite{little_decay_1967}. 
Although the low $\delta\Phi$ achieved at high temperature degrades the robustness of the memory with respect to flux-noise, it also allows to write the memory cell with smaller fluxes for a total cost of operation down to $\sim 10$~meV.

\section*{Discussion}
In summary, we have envisioned and demonstrated an original persistent Josephson phase-slip single memory cell which takes advantage of fluxoid quantization to codify two logic states in the topological index of the system, i.e., the parity of the superconducting winding number~\cite{strambini_-squipt_2016}. 
Differing from conventional superconducting loops~\cite{ilin_supercurrent-controlled_2021,murphy_nanoscale_2017,zhao_compact_2018}, here the  separation between the two topological states is provided by the large phase-slip barrier, which is unique to long superconducting JJs \cite{little_decay_1967,virtanen_spectral_2016}. 
Moreover, its operation mechanism is completely independent of the size or inductance of the superconducting loop thus allowing device miniaturization only limited by fabrication capabilities.
The memory exploits conventional superconductors thereby avoiding the use of complex ferromagnetic metals typical of present superconducting memories~\cite{ryazanov_magnetic_2012,gingrich_controllable_2016,baek_hybrid_2014,golod_single_2015,madden_phase_2018,vernik_magnetic_2013}.
Notably, the performances of the PSM are competing with state-of-the-art superconducting memories with an extremely low energy dissipation per bit operation ($\sim 10^{-24}$~J and $\sim 10^{-20} $~J for readout and write, respectively) 
and high operation speed (up to $\sim 30 $~ps and $\sim 1 $~ps for readout and write, respectively).  
Thanks to the topological protection, the PSM shows endurance, persistence, and high-temperature operation (up to $\sim 1 $~K), only limited by the Al critical temperature.
The use of vanadium~\cite{ligato_high_2017}
or niobium~\cite{jabdaraghi_low-temperature_2016}, 
therefore, could extend the memory operation above liquid He temperature, and further promote miniaturization thanks to the lower coherence length of these metals respect to Al.

In addition, our phase-based read-out scheme ensures protection against magnetic flux fluctuations, and provides ideal visibility in all the operation ranges. In fact, despite being intrinsically slower than conventional methods (high-speed lock-in amplifiers reach nowadays a clock frequency of  about $\sim 600$~MHz), the phase-based readout can be a valuable approach for the readout of phase-slip qubits. 
Furthermore, scalability to large arrays of PSM cells might be designed by taking advantage of the well known architectures employed for transition edge sensors, since both devices are based on a precise resistance measurement. In particular, frequency domain multiplexing or microwave resonators together with SQUID amplifiers \cite{ullom_review_2015} 
could be used for the selective read-out of each PSM composing the total memory.

Sneak currents can be avoided by employing strongly non-linear resistors between each single memory unit, such as superconductor/insulator/normal metal/insulator/superconductor Josephson junctions.
Integrating superconducting current feed lines in the ring\cite{enrico_-chip_2019} will allow to scale also the write procedure with the additional cost of wiring complexity.
Yet, the presence of independent write and read lines, with the former characterized by a low impedance, increases stability against perturbations of the read current and might simplify the integration of the PSM with existing superconducting logic elements including rapid single flux quantum\cite{golod_single_2015,zhao_compact_2018,ryazanov_magnetic_2012}, 
reciprocal quantum logic~\cite{Herr}, 
quantum flux parametrons~\cite{Hosoya}, 
Josephson field-effect transistors~\cite{Doh}, 
and gate-controlled cryotrons~\cite{de_simoni_metallic_2018,Paolucci1, Paolucci2}.
Yet, the strong topological protection and stability observed in the PSM make our approach promising in light of the implementation of phase-slip flux qubits \cite{mooij_phase-slip_2005,mooij_superconducting_2006} and quantum memories. 


\section*{Methods} \label{sec:Methods}

\subsection*{Device fabrication details.}  
\label{sec: Device fabrication details} 
The hybrid memory cells were realized by shadow-mask lithography technique. The suspended resist-mask was defined by electron-beam lithography (EBL) onto a SiO$\textsubscript{2}$ wafer. All metal-to-metal clean interfaces, and metal-to-oxide barriers were realized in an ultra-high vacuum (UHV) electron-beam evaporator (EBE) with a base pressure of 10$\textsuperscript{-11}$ Torr equipped with a tiltable sample holder suitable for multi-directional depositions. 
In order to obtain wire/ring transparent interfaces, which is crucial for the device operation, the use of the same material is strongly recommended\cite{ronzani_phase-driven_2017}. 
Therefore, the nanowire and the ring of the PSM were realized with aluminum. Furthermore, the Al film evaporation is relatively simple, and its high-quality native oxide allows the realization of good tunnel barriers through oxygen exposure at room temperature.
At first, 15 nm of Al$\textsubscript{0.98}$Mn$\textsubscript{0.02}$ were evaporated at an angle of -18$^\circ$ to realize the normal metal electrode. Subsequently, the sample was exposed to 60 mTorr of O$\textsubscript{2}$ for 5 min in order to form the thin insulating AlMnOx layer. Next, the sample holder was tilted to 10$^\circ$ for the deposition of 20 nm of Al realizing the SQUIPT nanowire (length ${L}$ = 400 nm, width $w=90$ nm and thickness $t=25$ nm) and the superconducting electrodes. Finally, a thicker layer of Al ($t_{R}=70$ nm) was evaporated at 0$^\circ$ to realize the superconducting loop of circumference $\sim7.6\;\mu$m, and average width $w_{R,ave}\simeq600$ nm. 

\subsection*{Magneto-electric characterization.} 
\label{sec:Magneto-electrical characterization}
The  magneto-electric characterization of the samples was performed at cryogenic temperatures in a $\textsuperscript{3}$He-$\textsuperscript{4}$He dilution refrigerator (Triton 200, Oxford Instruments) equipped with RC-filters of resistance $\sim$ 2k$\Omega$. 
The out-of-plane magnetic field was applied via a superconducting magnet driven by a low-noise current source (Series 2600, Keithley Instruments).
The DC measurements were performed in a two-wire voltage-bias configuration through a low-noise voltage DC source (GS200, Yokogawa) coupled with a room-temperature current preamplifier (Model 1211, DL Instruments) (see Fig. 1-c). 
The AC characterization was performed via a combination of DC bias and low-frequency lock-in technique. A DC bias voltage ($V$) was applied to the device. A current given by the sum of a DC and AC sinusoidal modulation energized the superconducting magnet. The read-out current oscillations induced by variation of $\Phi$, and the phase of the signal (with respect to the  flux oscillations) were recorded by a lock-in amplifier (SR830, Stanford Research Systems). Further details on the readout scheme can be found in the note 5 of the Supplementary Information.

\subsection*{Device parameters.} 
\label{sec:Device parameters}
Based on the device structure, we estimate the zero-temperature nanowire coherence length $\xi\textsubscript{w,0}$ = $\sqrt{\hbar D/\Delta \textsubscript{w,0}} \simeq 65$ nm, where $\hbar$ is the reduced Planck constant, ${D} \simeq {18}$ cm$\textsuperscript{2}$s$\textsuperscript{-1}$ is the diffusion coefficient, 
and $\Delta \textsubscript{w,0}\simeq$ 200 $\mu$eV is the zero-temperature gap in Al. 
The nanowire critical temperature is $T_{C,w}=\Delta \textsubscript{w,0}/1.764k_B\simeq1.31$ K, where $k_B$ is the Boltzmann constant. 
At low temperature, the ratio ${L}/ \xi\textsubscript{w,0}\simeq 6$ confirming the frame of the long JJ regime for the PSM.\cite{likharev_superconducting_1979}.
The single-valued CPR limit (achieved for $\xi_{w,short}\gtrsim L/3.5\sim 114$ nm) is reached at temperature $T_{short}=T_{C,w}(1-0.852^2\frac{\xi_{w,0}l}{\xi^2_{w,short}})\sim 1.29$~K~\cite{likharev_superconducting_1979}, where $l=3D/v_F\simeq 3$ nm is the nanowire mean free path, and  $v_F=2.03\times10^6$~m/s is the Fermi velocity of Al.

The kinetic inductance ($\mathcal{L}_K$) of a long JJ depends on the geometry and superconducting properties of the nanowire\cite{virtanen_spectral_2016}. 
In our case, at $25$  mK it takes the value $\mathcal{L}_K=\frac{R_N\hbar}{\pi\Delta \textsubscript{w}}\frac{1}{\tanh{\frac{\Delta \textsubscript{w}}{2k_B T}}}\simeq18$~pH~\cite{meservey_measurements_1969}. 
The nanowire normal-state resistance is given by $R_N=\frac{L}{wt\sigma}\simeq17\;\Omega$, where $\sigma=DN_fe^2 \simeq 1 \times10^7$ S m$^{-1}$ is the Al film conductance (with $N_f=2.15\times10^{47}$ J$^{-1}$m$^{-3}$ the density of states at the Fermi energy of Al). Analogously, the ring total inductance (including both the geometric and kinetic contributions) takes the value $\mathcal{L}_R \sim$ 1 pH~\cite{ronzani_phase-driven_2017}(with normal-state resistance $R_{R}\simeq1.4\;\Omega$). 
The contribution of the ring to the total inductance of the SQUIPT
yields a screening parameter $\beta=\mathcal{L}_R/\mathcal{L}_K \lesssim 0.1$.
The small $\beta$ cannot account for the hysteretic behavior of the PSM, which stems, differently, from the long-junction regime of the Josephson nanowire.
The writing/erasing time ($\tau_{W,E}$) is mainly due to the time required to polarize the SQUIPT with the external  flux. It is given by $\tau_{W,E}=\mathcal{L}_{SQUIPT}/R_{SQUIPT}\sim $ 1 ps, where $\mathcal{L}_{SQUIPT}=\mathcal{L}_{K}+\mathcal{L}_{R}$ and $R_{SQUIPT}=R_{N}+R_{R}$ are the total inductance and resistance of the SQUIPT, respectively. 
The read-out time ($\tau_{R}$) is predominantly limited by the characteristic time of the two tunnel barriers, $\tau_{R}=\tau_{t1}+\tau_{t2}\sim 30$ ps, where $\tau_{t1}=R_{t1}C_{t1}\sim 20$ ps is the characteristic time of the first tunnel junction, and $\tau_{t2}=R_{t2}C_{t2}\sim 10$ ps is the time constant of the second junction. The junctions capacitances ($C_{t1}\sim 0.3$ fF and $C_{t1}\sim 0.1$ fF) are estimated from the area and the typical specific capacitance of AlOx tunnel barriers $\sim 50$ fF/$\mu$m$^2$ 

\subsection*{Phase-slip rates}
Stochastic phase-slips are possible via quantum tunneling and thermal activation. They scale exponentially with the phase-slip barrier, the former with $-U/\Delta\textsubscript{w,0}$ while the latter with $-U/k_B T$. Both of them are small for $R_{\xi} < R_{q}$ (where $R_{\xi}=R_N \xi_{w}/L$),  as demonstrated in the following. 
The quantum phase-slip rate is~\cite{mooij_phase-slip_2005}:
\begin{equation}
    \Gamma_{qps} = \Omega_{qps} \exp{-0.3 \frac{R_q}{R_{\xi}}}, 
\end{equation}
where $\Omega_{qps} \simeq 0.85 \frac{\Delta\textsubscript{w}}{\hbar} \frac{L}{\xi_w} \sqrt{\frac{R_q}{R_{\xi}}} \simeq 75 $~THz is the quantum phase-slips attempt frequency.
With the parameters of our experiment we obtain the negligibly small $\Gamma_{qps} \sim 2\times 10^{-289}$~Hz.
Thermally activated phase-slips rate reads~\cite{arutyunov_superconductivity_2008}:
\begin{equation}
    \Gamma_{TAPS} = \Omega_{TAPS} \exp{-\frac{\delta F}{k_B T}}, 
\end{equation}
where $\delta F = 2.7 \frac{Tc-T}{T} U $ is the free energy difference of the potential barrier and 
$\Omega_{TAPS} \simeq 5.5 \frac{k_B T}{\hbar} \frac{L}{\xi_w} \sqrt{\frac{\delta F}{k_B T}} $ is the attempt frequency. In the temperature range of the experiment $ T<< Tc$, $\Gamma_{TAPS}$ is expected to be even smaller then $\Gamma_{qps}$. As an example, at $T= 100 $~mK the attempt frequency is $\Omega_{TAPS} \simeq 500$~THz and $\Gamma_{TAPS} \sim 10^{-474,257}$Hz. From these equations is possible to see that $\Gamma_{TAPS}$ is relevant only at temperature very close to Tc.

\section*{Data Availability}
The data that support the findings of this study are available from the corresponding author upon reasonable request.

\section*{Acknowledgements}
The authors acknowledge M. Cuoco and P. Virtanen for fruitful discussions.
N.L., E.S., and F.G. acknowledge partial financial support from the European Union’s Seventh Framework Programme (FP7/2007-2013)/ERC Grant No. 615187- COMANCHE.  N.L., E.S., and F.G. were partially supported by EU’s Horizon 2020 research and innovation program under Grant Agreement No. 800923 (SUPERTED). The work of F.P. was partially supported by the Tuscany Government (Grant No POR FSE 2014-2020) through the INFN-RT2 172800 project. 
The authors acknowledge the European Union (Grant No. 777222 ATTRACT) through the T-CONVERSE project. 

\section*{Author contributions} \label{sec:Author contributions}
E.S. and F.G. conceived the experiment. N.L. fabricated the samples with inputs from F.P.. N.L. and E.S. performed the measurements. N.L. analyzed the experimental data with inputs from E.S. and F.G.. 
All the authors discussed the results and their implications equally at all stages and wrote the manuscript.

\section*{Competing Interests} \label{sec: Additional information}
The authors declare no competing interests


\bibliographystyle{naturemag_NoURL}
\bibliography{BibElia.bib}

\clearpage
\onecolumngrid


\begin{center}
        \vspace*{0.5cm}
        \Large
        \textbf{Supplementary Information}
        \vspace{0.5cm}
\end{center}

\renewcommand{\thefigure}{S\arabic{figure}}
\renewcommand{\theequation}{S\arabic{equation}}
\setcounter{figure}{0}

\section{Hysteresis in the current vs voltage characteristics}

To test the PSM transport properties and highlight the hysteresis in its magneto-resistance, we electrically characterized the device at $T=25$ mK. Figure S1 shows the current vs voltage characteristics ($\textit{I}(\textit{V})$) of a typical PSM measured at $\Phi= 0.45\Phi\textsubscript{0}$ (blue curve),  $\Phi= 0.5\Phi\textsubscript{0}$ (red curve) and $\Phi= 0.54\Phi\textsubscript{0}$ (green curve) for positive (left) and negative (right) sweeps of the magnetic flux. 
The tunnel $I$($V$) characteristics reveal a magnetic-flux-induced modulation of the superconducting gap of the Al nanowire ($\Delta \textsubscript{w}$), and hysteric behavior with $\Phi$, showing a maximum reduction of $\Delta \textsubscript{w}$ at $\Phi$ = $0.54\Phi\textsubscript{0}$ and $\Phi = 0.45\Phi\textsubscript{0}$ for the forward and backward traces, respectively.

\begin{figure}[ht!]
        \centering
  \includegraphics[width=0.5\linewidth]{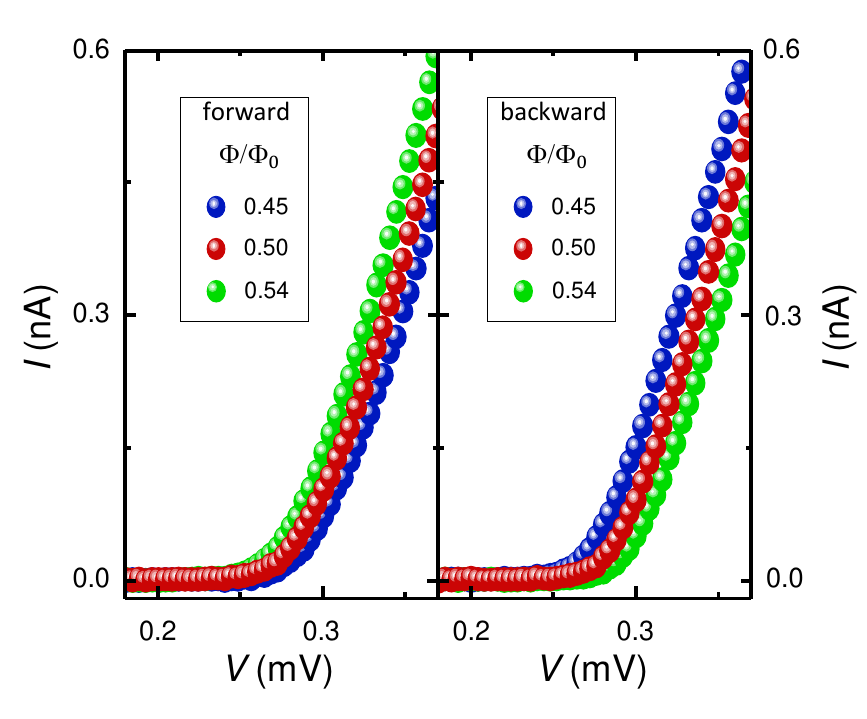}
  \label{S1}
        \caption{\textbf{Current ($I$) vs Voltage ($V$) curves measured at different values of magnetic flux ($\Phi$).} Left panel: $I$($V$) recorded while increasing $\Phi$ (forward). Right panel: $I$($V$) recorded while decreasing $\Phi$ (backward). All data were recorded at $\textit{T} = 25$ mK.}
\end{figure}

\clearpage
\section{PSM in continuous read-out configuration for several applied voltage biases}
In order to find the optimal operating parameters for the PSM, we performed the writing/erasing operations in the continuous read-mode, i.e., when a bias voltage ($\textit{V}$) is permanently applied, varying the values of $V$ from 200 $\mu$V to 600 $\mu$V, as shown in Fig. S2 at $T=25$ mK. The bias flux is set just above the crossing-point of the hysteresis, namely at $0.54\Phi_0$.
The memory can be written or erased by applying a flux pulse down to $\Phi\textsubscript{W}=0.33\Phi\textsubscript{0}$ or up to $\Phi\textsubscript{E}=0.75\Phi\textsubscript{0}$, respectively.

The PSM shows the typical behavior of a memory cell with distinct current values for $[0]$ and $[1]$ states for $V\geq300;\mu$V. By increasing $V$ the visibility between the two stored states increases until saturating for the largest biases. This trend is due to local overheating in the weak link induced by the quasiparticle current flowing through the probing junction, which increases $\xi\textsubscript{w}$  thereby deviating the CPR towards the single-valued non-hysteretical form. Finally, we stress that the memory works properly for many bias cycles, confirming the \textit{endurance} of the PSM cell.

\begin{figure}[h!]
        \centering
  \includegraphics[width=0.9\linewidth]{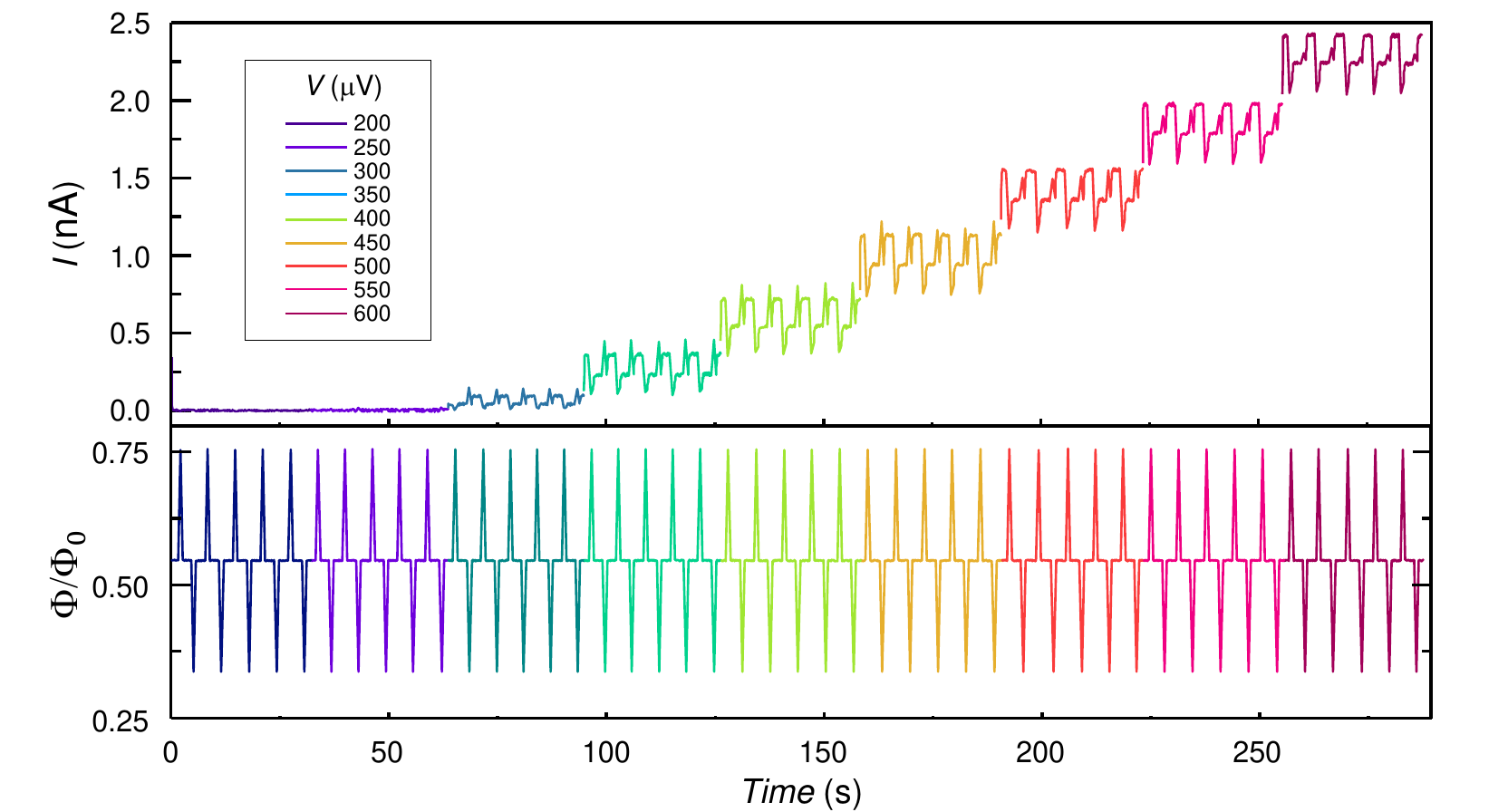}
        \caption{\textbf{Writing/erasing operations in the continuous read-out mode performed for different values of permanently applied voltage bias   ($\textit{V}$).} Bottom: evolution of the magnetic flux composed by a bias at ($\Phi\textsubscript{R}=0.54\Phi\textsubscript{0}$) and pulses to write ($\Phi\textsubscript{W}=0.33\Phi\textsubscript{0}$) and erase ($\Phi\textsubscript{E}=0.75\Phi\textsubscript{0}$) the memory state. Top: evolution of the read-out current ($\textit{I}$) measured for different values of voltage bias ($\textit{V}$). The higher (lower) current value $\textit{I}\textsubscript{[1]}$ ($\textit{I}\textsubscript{[0]}$) acquired for a fixed $\textit{V}$ corresponds to logic state [1] ([0]). $\textit{I}\textsubscript{[1]}$ and $\textit{I}\textsubscript{[0]}$ are negligible for $\textit{V}< 300 \mu$V, therefore the PSM cannot be biased in that voltage range. All data were acquired at $\textit{T} = 25$ mK.}
\end{figure}

\clearpage
\section{Non-volatility of the PSM measured at different voltage biases}

Here, we study the \textit{non-volatility} of the PSM. Since the quiescent magnetic flux could be provided by a metallic ferromagnet buried in the isolating substrate or by a ferromagnetic insulator dielectric, the only power source relevant for the PSM is the voltage bias ($V$). Figure S3 shows the non-volatility of the device measured for several values of $V$.
Temporarily removing the voltage bias has no effect on the stored data. In fact, the reading voltage is set to $V$ only during the readout operation. In addition, the writing, erasing and reading operations have been performed several times without any sizeable outcome change.

\begin{figure*}[h!]
       \centering
  \includegraphics[width=\linewidth]{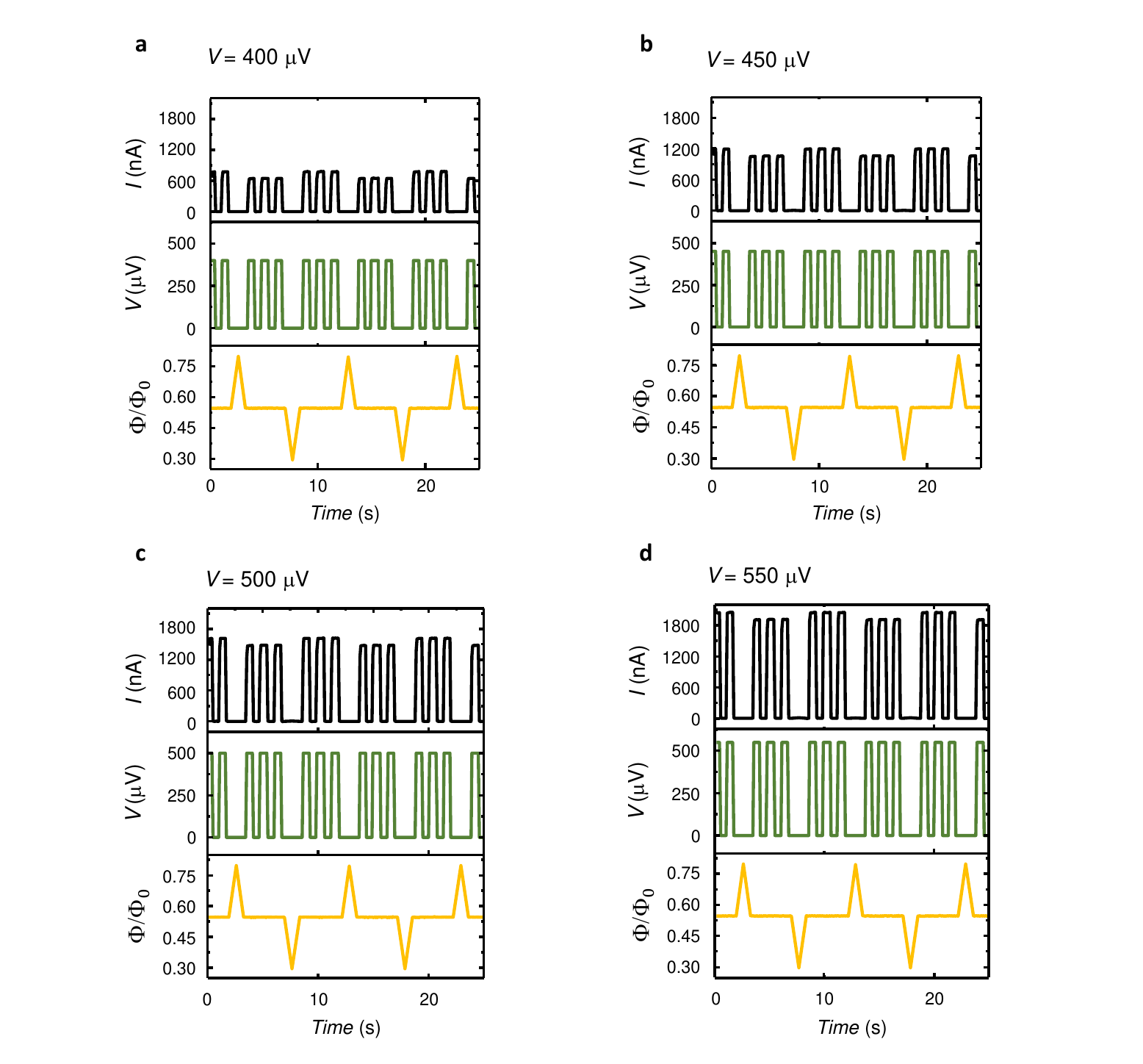}
        \caption{\textbf{Writing/erasing operations performed for different values of a pulsed readout voltage bias ($\textit{V}$).} $\textit{V}$ ranges form: $400\;\mu$V (\textbf{a}), $450\;\mu$V (\textbf{b}), $500\;\mu$V (\textbf{c}) and $550\;\mu$V (\textbf{d}). 
        For each panel is shown the temporal evolution of the magnetic flux applied to the PSM (bottom) together with the bias voltage pulses (center) and the resulting tunnel current (top) used for the state readout. The flux was biased at $\Phi\textsubscript{B}=0.54\Phi\textsubscript{0}$ while flux pulses of $\Phi\textsubscript{W}=0.33$ $\Phi\textsubscript{0}$ and $\Phi\textsubscript{E}=0.75$ $\Phi\textsubscript{0}$ were used for write and erase operations, respectively. 
        The permanency of the memory state before and after the application of the readout bias demonstrate the persistency and non-volatility of the PSM. Note that the PSM works properly during the operation of the device for a wide range of applied voltage biases. All measurements were performed at $\textit{T} = 25$ mK.} 
        \label{delta}
\end{figure*}

\clearpage
\section{PSM operation at the degenerate flux $\Phi_{B}=0.5\Phi_0$}

This section is devoted to the study of the properties of the PSM when operated at the degenerate flux $\Phi\textsubscript{B}=0.5\Phi_0$, where the two current branches are expected to show the same value. To this end, we bias the device with a constant voltage ($V$) while a sinusoidal oscillation of the magnetic flux is superimposed on $\Phi\textsubscript{B}=0.5\Phi_0$. Figure S4 confirms that the average values of $\textit{I}\textsubscript{[0]}$ and $\textit{I}\textsubscript{[1]}$ are indistinguishable. Conversely, the readout current oscillates in reversed phase depending on the stored state. Therefore, the information of the memory state is stored in the phase of the current signal, thus enabling the design of a phase-dependent readout.

\begin{figure}[h!!]
        \centering
  \includegraphics[width=0.9\linewidth]{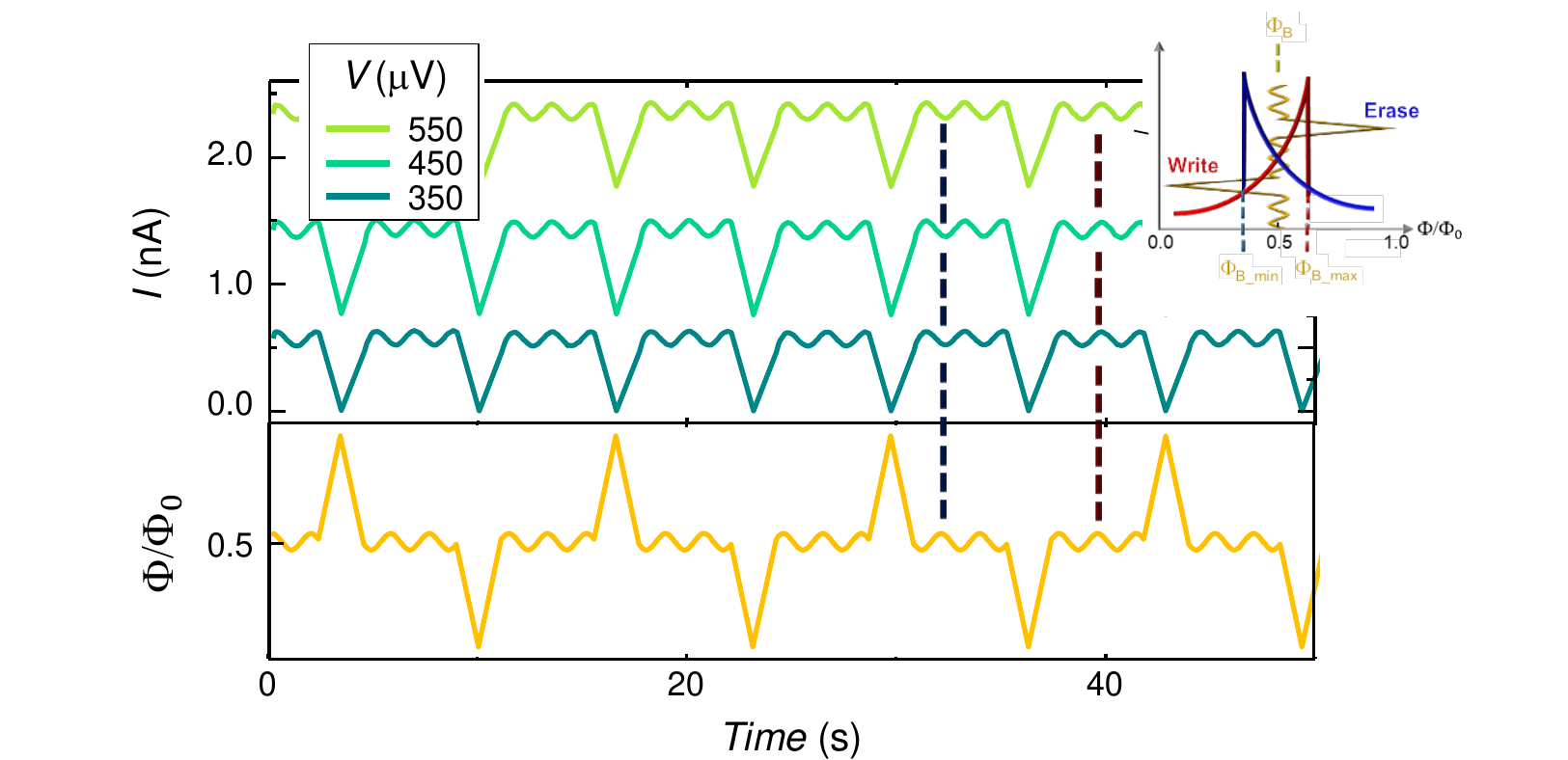}
        \caption{\textbf{Writing/erasing operations performed at the degenerate point $\Phi\textsubscript{B}=0.5\Phi_0$.} Bottom: Temporal evolution of the total flux applied to the PSM and composed by a bias $\Phi\textsubscript{B}=0.5\Phi\textsubscript{0}$ superimposed to a low frequency sinusoidal fluctuation with amplitude $\Phi\textsubscript{AC}=\pm0.04\Phi\textsubscript{0}$. Write ($\Phi\textsubscript{W}=0.30\Phi\textsubscript{0}$) and erase ($\Phi\textsubscript{E}=0.80\Phi\textsubscript{0}$) pulses are applied to switch the memory state. Top: read-out current acquired at different voltage bias ($\textit{V}$). Notice that at the degeneracy point the average values of $\textit{I}\textsubscript{[0]}$ and $\textit{I}\textsubscript{[1]}$ are indistinguishable while information of the memory state is stored in the phase of the signal. $\textit{I}\textsubscript{[0]}$ and $\textit{I}\textsubscript{[1]}$ show a $180^{\circ}$ phase shift.Inset: scheme of operation in the $I(\Phi)$ diagram. All measurements were performed at $\textit{T} = 25$ mK.
} 
\end{figure}

\clearpage
\section{Phase-dependent read-out scheme}

The phase-dependent read-out can be realized by means of the experimental setup shown in Fig. S5a. The total magnetic flux is the sum of a constant component $\Phi\textsubscript{B}=0.5\Phi_0$ (due to the current $I_B$ flowing in the superconducting magnet) and a small sinusoidal component $\Phi_{AC}$ (due to the current $I_{AC}$). The PSM is biased with a constant voltage $\textit{V}$. The phase of the output current ($\phi$) is measured with a lock-in amplifier with respect to the $I_{AC}$ oscillations. The results are summarized in Fig. S5b. The readout current for state $[1]$ oscillates in phase with the magnetic flux oscillations, while state $[0]$ shows a counter phase (180$^\circ$) fluctuation. We note that the device unequivocally shows the typical behavior of a memory cell upon many erasing/writing cycles even for the phase-dependent readout.

\begin{figure}[h!]
        \centering
  \includegraphics[width=0.9\linewidth]{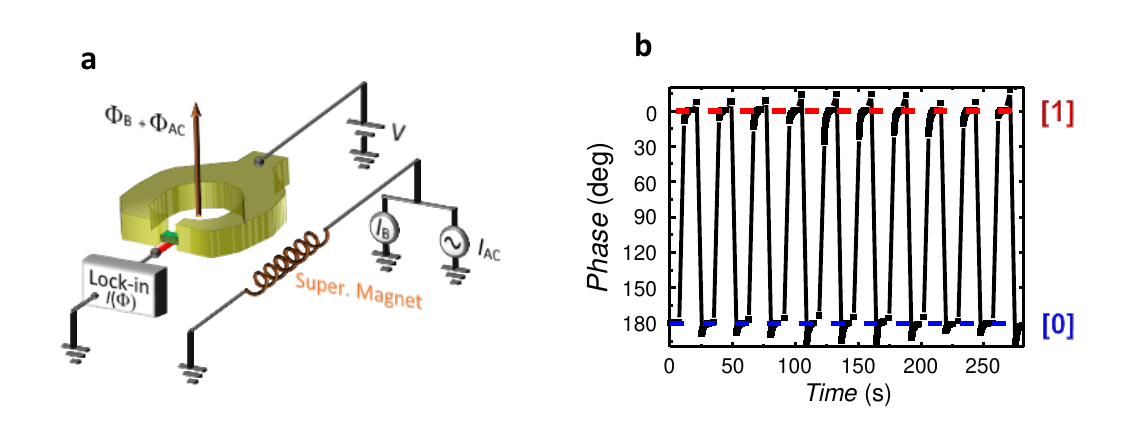}
        \caption{\textbf{Read-out of the PSM by lock-in phase measurements} \textbf{a} Scheme of the PSM, where the Al ring (yellow), the Al nanowire (green) and the normal metal tunnel probe (red) are presented together with the complete electrical set-up. The PSM is biased with a DC constant voltage $\textit{V}$= $400\;\mu$V. The magnetic flux piercing the superconducting ring is $\Phi_B+\Phi_{AC}$, where $\Phi_B=0.5\Phi_{0}$ is the flux corresponding to the crossing point of the two current branches and $\Phi_{AC}$ is a small sinusoidal component superimpose with a lock-in source. The variations of the read-out current with the flux oscillations are recorded with standard lock-in technique. \textbf{b} Time dependence of the phase of the readout current while writing and erasing operations are performed.
        $I_{[1]}$ oscillates in phase with the magnetic flux (phase=0), while $I_{[0]}$ has the opposite dependence (phase=180$^\circ$) allowing the acquisition of the two distinguishable signals of phase. All data were recorded at $\textit{T} = 25$ mK.}   
	\end{figure}
	
\end{document}